# Dust surface potential for the dusty plasma with negative ions and with a three-parameter non-Maxwell velocity distribution

Guoxue Yao and Jiulin Du

*Department of Physics, School of Science, Tianjin University - Tianjin 300072, China*

**Abstract** – We study the dust surface potential for the complex dusty plasma with negative ions and with a three-parameter non-Maxwell velocity distribution. The plasma contains electrons, positive ions, negative ions, and negatively charged dust particles. By using the current equilibrium condition, we derive the relationship between the normalized dust surface potential and the dusty plasma parameters such as the normalized dust number density, the temperature ratio of negative ions to electrons, the density ratio of negative ions to positive ions, and the charge number of negative ions. The numerical analyses show that the relationship depends evidently on the three parameters in the non-Maxwell distribution when the dust surface potential is relatively smaller, but with increase of the potential, such dependence will weaken soon. The dust surface potential is negative and increases monotonously with increase of the dust density, and for the complex dusty plasma with the three-parameter non-Maxwell distribution, it is generally greater than that in the same plasma with the kappa-distribution and the Maxwellian distribution.

**Introduction.** – Dust particles are some of solid grain matter common in the universe, while gaseous matter in astrophysical and space systems is often in the fully ionized or partially ionized plasma state. In this way, the plasma as well as the dust particles immersed in it constitutes a complex dusty plasma system. It is complex system because, in addition to the conventional electron-ion plasma, additional components such as neutral gas molecules, negative ions, and dust particles of large mass (relative to ions) or nebula clusters are added to the plasma system. Dusty plasmas exist in different parts of the cosmic environments such as interstellar medium, planetary rings, phobos-dust rings, comet tails and interstellar molecular clouds. At the same time, they also exist in the fusion reactors and the plasma processing industrial installations [1-4]. Plasma containing dust particles can be regarded as dusts in plasma or dusty plasma, which depends on some characteristic lengths such as the dusty particle radius $r_d$ and the average distance $a$ between particles (related to dust number density $n_d$, $n_d a^3 \sim 1$). If Debye radius is $\lambda_D$ and when there is $r_d \ll a < \lambda_D$, that is, charged dust particles participate in the collective behavior, corresponding to the dusty plasma. When there is $r_d \ll \lambda_D < a$, that is, charged dust particles are considered as collection of isolated dust grains, corresponding to dusts in plasma [5].

Now we consider the case of dust particles as large-mass charged particles, similar to positive or negative ions. Dust particles may be metallic, dielectric or ice (graphite, silicate, magnetite, and amorphous carbon, etc.)[5], so their sizes are usually different. The presence of charged dust particles not only modifies the existing low frequency waves, such as ion acoustic waves, low hybrid waves and ion-acoustic shocks, but also introduces new types of low-frequency waves related to the dust particles, such as dust acoustic shocks, dust ion-acoustic shocks, etc [5]. Therefore, the dusty plasma models have been verified and studied in experiments and also in theory in recent years [6-10]. Rao *et al* considered a collision-free fluid model and studied the dust acoustic wave [11]. Melandso studied the complex plasma lattice in a strongly coupled system and proposed the dust lattice wave [12]. Shukla and Silin found a new low-frequency electrostatic wave and a new dust-acoustic wave for the complex plasmas in planetary rings [13]. The basic components of dusty plasma are electrons, ions, neutral particles and dust particles. The biggest difference between dusty plasma and ordinary plasma is the charge and the discharge of dust particles, which thus results in many different properties of the dusty plasma from the ordinary plasma.

The most basic property of dusty plasma is the charge of dust particles. The charge on dust particles is not inherent, but it is from the plasma. There are many mechanisms for the dust particle charging are known, such as the orbit-limited electrostatic probe model[14], the photoelectron emission and secondary electron emission mechanism[5]etc, which were recently studied for the dusty plasma with power-law distributions (see[15,16] and the references therein). In this paper, we only consider the case where the dust particles are negatively charged.

Barkan et al were the first to study the charging process and the surface potential of dust particles in the dusty plasma with Maxwell-distributed electrons and positive ions [17]. Mamun and Shukla studied the charging of dust particles in the Maxwell -distributed dusty plasma with negative ions [18]. It was found that negative ions are an additional component in the experimental and spatial plasmas [19-23]. Negative ions, as the original source of the aggregation reaction, usually play an important role in the formation of atomic groups [24]. The role of negative ions in the characteristics of plasma is mainly reflected in the control of electrode potential in etching and deposition plasma [19,20]. Abid et al studied the surface potential energy in the dusty plasma with negative ions, highlighting the role of negative ions [25-28], where the plasma follows the kappa-distribution, a non-Maxwell velocity distribution.

When some or all of the plasma particles move faster than the thermal velocity, traditional Maxwell-distribution can no longer describe the statistical behavior of the superheated particles, so many non-Maxwell distributions are proposed and studied for complex plasmas. In 1968, Vasyliunas analyzed the energy spectrum of electrons in the space plasma magnetic shoe and introduced a high-energy electron velocity kappa-distribution function [29]. Cairns et al also studied a non-thermal velocity distribution function [30]. Abid et al used a combined two-parameter Vasyliunas-Cairns distribution [26] and a generalized Cairns-Tsallis distribution [27] to study the surface potential of the dust plasma containing negative ions. Ali et al also used a two-parameter distribution to study the dust plasma charging process and the effect of negative ions on the surface potential of dust particles [28]. It was found that the streaming velocity, the number density and the other parameters of negative ions have significant effect on the surface potential of dust particles in complex plasmas. Recently, a new universal three-parameter non-Maxwell velocity distribution was introduced [31], consisting of Cairns distribution and the two-parameter distribution [30,32], which is important for understanding many linear and nonlinear phenomena in the space plasmas, such as solar wind[33], magnets[34] and aurora belts[33,35]. In this work, we study the dust charging model of the complex dusty plasma with the three-parameter non-Maxwell distribution, and derive the property of the currents of electrons and ions (positive ions/negative ions) as well as the dust surface potential.

The layout of this paper is arranged as follows. In section 2, we introduce the three-parameter non-Maxwell velocity distribution in complex dusty plasma. In section 3, we derive the equation for the dust surface potential and the other plasma quantities such as the dust number density etc. In section 4, we make the numerical analyses, and finally in section 5, we give the conclusion.

**The three-parameter non-maxwell velocity distribution.** − Recently, a new and universal three-parameter non-Maxwell velocity distribution for complex dusty plasmas was introduced [31], which generally contained the most extensive velocity distributions of the complex plasmas, in certain cases of the parameters, such as the two-parameter $(r, q)$ distribution, the Vasyliunas-Cairns distribution, the power-law $q$-distribution in nonextensive statistics, the Cairns-distribution, the kappa-distribution, and also the Maxwellian distribution etc. The three-parameter non-Maxwell velocity distribution function can be written for the $j$th plasma component [31] as

$$f_j(V_j) = Y_j \left(1 + \alpha \frac{V_j^4}{V_{Tj}^4}\right)\left[1 + \frac{1}{q-1}\left(\frac{V_j^2}{X_{r,q}V_{Tj}^2}\right)^{r+1}\right]^{-q}, \tag{1}$$

where we denote $Y_j = \left(\frac{3N_j}{4\pi V_{Tj}^3}\right)\frac{\rho_{\alpha,r,q}}{X_{r,q}^{3/2}}$ (see Appendix (A.1) and (A.2)), with

$$\rho_{\alpha,r,q} = \frac{\Gamma(q)}{\left(1+9\eta_{r,q}\alpha\right)(q-1)^{\frac{3}{2(1+r)}}\Gamma\left(q-\frac{3}{2(1+r)}\right)\Gamma\left(1+\frac{3}{2(1+r)}\right)},$$

$$\eta_{r,q} = \frac{\Gamma\left(q-\frac{3}{2(1+r)}\right)\Gamma\left(\frac{3}{2(1+r)}\right)\Gamma\left(q-\frac{7}{2(1+r)}\right)\Gamma\left(\frac{7}{2(1+r)}\right)}{\Gamma^2\left(q-\frac{5}{2(1+r)}\right)\Gamma^2\left(\frac{5}{2(1+r)}\right)},$$

$$X_{r,q} = \frac{3\Gamma\left(\frac{3}{2(1+r)}\right)\Gamma\left(q-\frac{3}{2(1+r)}\right)}{(q-1)^{\frac{1}{r+1}}\Gamma\left(q-\frac{5}{2(1+r)}\right)\Gamma\left(\frac{5}{2(1+r)}\right)}.$$

In this velocity distribution function (1), $Y_j$ is the normalized coefficient, the three spectral parameters, $\alpha$, $r$ and $q$, describe the high-energy particles at the shoulder of the speed distribution curve, the high-energy particles on the broad shoulder, the superthermal and width of the tail of the velocity distribution curve, respectively; $V_{Tj} = (T_j/m_j)^{1/2}$ is the thermal speed with temperature $T_j$ and mass $m_j$ of the unit energy for $j$th component of particles in the dusty plasma, $N_j$ is the number density, the subscript $j = e$, $i$ and $n$ denotes the electron, positive ions and negative ions, respectively. The three parameters need to satisfy $q >1$, $\alpha >0$ and $q(1+r) >5/2$.

The three-parameter non-Maxwell distribution (1) has the following properties: when we take $\alpha = 0$, it reduces to the two-parameter $(r, q)$ distribution function; when we take $\alpha=0$ and $q=\kappa +1$, it reduces to the Vasyliunas-Cairns distribution; when we take $\alpha$ =0 and $r$ =0, it reduces to the $q$-distribution in nonextensive statistics; when we take $\alpha$ =0 and $q\to\infty$, it reduces to the Cairns distribution; when we take $\alpha$=0, $r$=0 and $q=\kappa+1$, it reduces to the famous kappa-distribution; and finally when we take $\alpha$=0, $r$=0 and $q\to\infty$, it reduces to the Maxwellian distribution in traditional statistics.

**The dust surface potential in the dusty plasma following the three-parameter non-maxwell distribution.** − We assume that spherical dust particles are immersed in the magnetic field-free plasma which contains electrons, positive ions, and additional negatively charged ions, and the plasma follows the three-parameter velocity distribution (1). We consider the plasma particle from a distance of infinity to approach the dust particle with a radius $r_d$ and a charge $Q_d$. When a charged particle enters the Debye ball, it is affected by dust particle, and due to electrostatic force among the dust grains and the plasma species bring about a negative charge on the dust grains. The dust particles are negatively charged, so that the electrostatic field established around them repels electrons and negative ions but attracts positive ions. With the negative potential increase at the surface of dust grains, the ion current approaching to the surface increases gradually but the electron current and negative current decrease. Finally, the electrons and negative ions that hit the dust particles' surface are equal to the ion current, and they achieve the electrostatic balance, satisfying the current equilibrium:

$$\partial Q_d / \partial t = \sum_j I_j = I_e + I_i + I_n = 0, \qquad (2)$$

where $I_j$ is the current associated with the $j$th component of the dusty plasma, i.e., $j=e, i, n$ for the electrons, the ions and the negative ions respectively. Based on the orbit motion limit theory [5], the current of $j$th component in the dusty plasma can be written as

$$I_j = 4\pi Q_j \int_{V_j^{\min}}^{V_j^{\max}} V_j^3 \sigma_j^d f_j(V_j) dV_j, \qquad (3)$$

where $f_j(V_j)$ is the velocity distribution function, $Q_j$ is the charge and $\sigma_j^d$ is the collision cross section for the charging collisions between the dust grains and the electrons, the ions or the negative ions, given [5] by

$$\sigma_j^d = \pi r_d^2 \left(1 - \frac{2Q_j \phi_d}{m_j V_j^2}\right), \qquad (4)$$

where $m_j$ is the mass of the particles, and the dust surface potential, $\phi_d$ (a potential relative to average potential of the dusty plasma), is related to its charge $Q_d$ by $\phi_d = Q_d/r_d$. The cross section $\sigma_j^d$ must be positive so that the collisions can take place, which requires the velocity to be $V_j \geq \sqrt{2Q_j \phi_d / m_j}$ if $Q_j \phi_d > 0$. Because $\phi_d$ is negative, from Eq.(3) we have that

$$I_e = -4\pi e \int_{\sqrt{2Q_e \phi_d / m_e}}^{\infty} V_e^3 \sigma_e^d f_e(V_e) dV_e, \qquad (5)$$

$$I_i = 4\pi Z_i e \int_0^{\infty} V_i^3 \sigma_i^d f_i(V_i) dV_i, \qquad (6)$$

$$I_n = -4\pi Z_n e \int_{\sqrt{2Q_n \phi_d / m_n}}^{\infty} V_n^3 \sigma_n^d f_n(V_n) dV_n, \qquad (7)$$

where we have that $Q_e = -e$, $Q_i = Z_i e$ and $Q_n = -Z_n e$ for the electron, the ion and the negative ion respectively.

After the integrations in Eqs. (5)-(7) are completed (see Appendix), we derive that

$$I_e = -6\pi e N_e r_d^2 \vartheta_e \frac{(q-1)^q \rho_{\alpha,r,q} U^2}{X_{r,q}^2} \left(\frac{X_{r,q}}{-2U}\right)^{q(1+r)} \left[4\alpha U^2 \left(\frac{H_1}{q+qr-4} - \frac{H_2}{q+qr-3}\right)\right.$$
$$\left. + \frac{H_3}{q+qr-2} - \frac{\Gamma\left(q-\frac{1}{1+r}\right) H_4}{(1+r)\Gamma\left(q+\frac{r}{1+r}\right)}\right], \qquad (8)$$

$$I_i = 3\pi r_d^2 \vartheta_i N_i Z_i e \left[A_{\alpha,r,q} \alpha X_{r,q}^2 \left(\frac{1}{8} - \frac{Z_i e \phi_d}{3 m_i \vartheta_i^2} B_{\alpha,r,q}\right) + C_{\alpha,r,q} \left(1 - \frac{2Z_i e \phi_d}{m_i \vartheta_i^2} D_{\alpha,r,q}\right)\right], \qquad (9)$$

$$I_n = -6\pi Z_n^2 e N_n r_d^2 \vartheta_n \left(\frac{\gamma X_{r,q}}{-2Z_n U}\right)^{q(1+r)} \frac{\rho_{\alpha,r,q} U^2 (q-1)^q}{\gamma^2 X_{r,q}^2} \left[\frac{4\alpha Z_n^2 U^2}{\gamma^2} \left(\frac{h_1}{q+qr-4} - \frac{h_2}{q+qr-3}\right)\right.$$
$$\left. + \frac{h_3}{q+qr-2} - \frac{h_4 \Gamma\left(q-\frac{1}{1+r}\right)}{(1+r)\Gamma\left(\frac{r}{1+r}+q\right)}\right] = 0, \qquad (10)$$

where $\vartheta_j = X_{r,q}^{1/2} V_{Tj}$ is the effective thermal speed, $U = e\phi_d / T_e$ is called the normalized dust surface potential. In Eq. (8) and Eq.(10), the eight abbreviations $H_1 \sim H_4$ and $h_1 \sim h_4$ denote eight hyper-geometric functions [36] respectively, given by

$$H_1 = {}_2F_1\left(q,\ q-\frac{4}{1+r},\ 1+q-\frac{4}{1+r},\ -\left(\frac{X_{r,q}}{-2U}\right)^{1+r}(q-1)\right),$$

$$H_2 = {}_2F_1\left(q,\ q-\frac{3}{1+r},\ 1+q-\frac{3}{1+r},\ -\left(\frac{X_{r,q}}{-2U}\right)^{1+r}(q-1)\right),$$

$$H_3 = {}_2F_1\left(q,\ q-\frac{2}{1+r},\ 1+q-\frac{2}{1+r},\ -\left(\frac{X_{r,q}}{-2U}\right)^{1+r}(q-1)\right),$$

$$H_4 = {}_2F_1\left(q,\ q-\frac{1}{1+r},\ 1+q-\frac{1}{1+r},\ -\left(\frac{X_{r,q}}{-2U}\right)^{1+r}(q-1)\right);$$

$$h_1 = {}_2F_1\left(q,\ q-\frac{4}{1+r},\ 1+q-\frac{4}{1+r},\ -\left(\frac{X_{r,q}\gamma}{-2Z_nU}\right)^{1+r}(q-1)\right),$$

$$h_2 = {}_2F_1\left(q,\ q-\frac{3}{1+r},\ 1+q-\frac{3}{1+r},\ -\left(\frac{X_{r,q}\gamma}{-2Z_nU}\right)^{1+r}(q-1)\right),$$

$$h_3 = {}_2F_1\left(q,\ q-\frac{2}{1+r},\ 1+q-\frac{2}{1+r},\ -\left(\frac{X_{r,q}\gamma}{-2Z_nU}\right)^{1+r}(q-1)\right),$$

$$h_4 = {}_2F_1\left(q,\ q-\frac{1}{1+r},\ 1+q-\frac{1}{1+r},\ -\left(\frac{X_{r,q}\gamma}{-2Z_nU}\right)^{1+r}(q-1)\right).$$

In Eq. (9), the abbreviations denote that

$$A_{\alpha,r,q} = \frac{(q-1)^{\frac{5}{2(r+1)}}\,\Gamma\left(1+\frac{4}{1+r}\right)\Gamma\left(q-\frac{4}{1+r}\right)}{(1+9\eta_{r,q}\alpha)\Gamma\left(q-\frac{3}{2(1+r)}\right)\Gamma\left(1+\frac{3}{2(1+r)}\right)},$$

$$B_{\alpha,r,q} = \frac{\Gamma\left(1+\frac{3}{1+r}\right)\Gamma\left(q-\frac{3}{1+r}\right)}{(q-1)^{\frac{1}{1+r}}\,\Gamma\left(1+\frac{4}{1+r}\right)\Gamma\left(q-\frac{4}{1+r}\right)},$$

$$C_{\alpha,r,q} = \frac{(q-1)^{\frac{1}{2(r+1)}}\,\Gamma\left(1+\frac{2}{1+r}\right)\Gamma\left(q-\frac{2}{1+r}\right)}{4(1+9\eta_{r,q}\alpha)\Gamma\left(q-\frac{3}{2(1+r)}\right)\Gamma\left(1+\frac{3}{2(1+r)}\right)},$$

$$D_{\alpha,r,q} = \frac{2\Gamma\left(1+\frac{1}{1+r}\right)\Gamma\left(q-\frac{1}{1+r}\right)}{(q-1)^{\frac{1}{1+r}}\,\Gamma\left(1+\frac{2}{1+r}\right)\Gamma\left(q-\frac{2}{1+r}\right)}.$$

We substitute Eqs. (8)-(10) into the current balance condition Eq.(3) and use the quasi-neutral condition,

$$N_e + Z_n N_n - Z_i N_i = Q_d N_d / e, \tag{11}$$

where $N_e$, $N_n$, $N_i$ and $N_d$ are the number density of electrons, negative ions, ions and dust particles respectively. And then we can derive the equation for the normalized dust surface potential $U$ and the plasma quantities such as the normalized dust number density, the temperature ratio of negative ions to electrons, the density ratio of negative ions to positive ions, and the charge number of negative ions etc.,

$$-2\mu U^2\left(1-\frac{Z_n}{Z_i}\varepsilon + Z_i PU\right)\left(\frac{X_{r,q}}{-2U}\right)^{q(1+r)}\frac{\rho_{\alpha,r,q}(q-1)^q}{X_{r,q}^2}\left[4\alpha U^2\left(\frac{H_1}{q-4+qr}-\frac{H_2}{q-3+qr}\right)\right.$$

$$+\frac{H_3}{q-2+qr}-\frac{H_4\Gamma\left(q-\frac{1}{1+r}\right)}{(1+r)\Gamma\left(q+\frac{r}{1+r}\right)}\left] + \sqrt{\sigma}\left[\alpha A_{\alpha,r,q}\left(\frac{X_{r,q}^2}{8}-\frac{Z_i UX_{r,q}}{3\sigma}B_{\alpha,r,q}\right)+C_{\alpha,r,q}\left(1-\frac{2Z_i UM_{\alpha,r,q}}{\sigma}\right)\right]$$

$$-2U^2\varepsilon\beta\sqrt{\gamma}\,\frac{Z_n^3}{Z_i}\left(\frac{\gamma X_{r,q}}{-2Z_nU}\right)^{q(1+r)}\frac{\rho_{\alpha,r,q}(q-1)^q}{\gamma^2 X_{r,q}^2}\left[\frac{4\alpha Z_n^2 U^2}{\gamma^2}\left(\frac{h_1}{q-qr+4}-\frac{h_2}{q-qr+3}\right)\right.$$

$$\left.+\frac{h_3}{q+qr-2}-\frac{h_4\Gamma\left(q-\frac{1}{1+r}\right)}{(1+r)\Gamma\left(\frac{r}{1+r}+q\right)}\right] = 0 \tag{12}$$

where the plasma quantities [16,19] are $\sigma = T_i/T_e$, $\mu = (m_i/m_e)^{1/2}$, $\gamma = T_n/T_e$, $\varepsilon = N_n/N_i$, $\beta = (m_i/m_n)^{1/2}$, $M_{\alpha,r,q} = D_{\alpha,r,q}/X_{r,q}$, and $P = 4\pi\lambda_0^2 N_d r_d$ with the characteristic length, $\lambda_0 = \sqrt{T_e/4\pi Z_i^2 e^2 N_i^2}$. $P$ is called the normalized number density of dust particles. Eq. (12) determines a relation between the normalized dust surface potential and the other plasma quantities above in the dusty plasma with the three-parameter non-Maxwell velocity distribution.

In the above case, Eq.(12) can determine a relationship between $U$ and $P$ in the dusty plasma in the two different cases of negative ions and with the three-parameter non-Maxwell velocity distribution, which is equivalent to the relationship between the dust surface potential $\phi_d$ and the number density $n_d$ of dust particles. The dependence of the relationship between $U$ and $P$ on the non-Maxwell velocity distribution can be studied in accordance with different values of the three parameters $\alpha$, $r$ and $q$.

If we take the three parameters as $\alpha = r = 0$ and $q = \kappa + 1$, Eq. (12) becomes the case for the same dusty plasma with the $\kappa$-distribution. Namely, Eq.(12) becomes

$$-\mu\left(1-\frac{Z_n}{Z_i}\varepsilon + Z_i PU\right)\frac{2U^2\kappa^{\kappa+1}}{X_{0,\kappa+1}^2}\left(\frac{X_{0,\kappa+1}}{-2U}\right)^{\kappa+1}\frac{\rho_{0,0,\kappa+1}}{C_{0,0,\kappa+1}}\left\{\frac{1}{\kappa-1}\,_2F_1\left(\kappa+1,\kappa-1,\kappa,\left(\frac{X_{0,\kappa+1}}{2U}\right)\kappa\right)\right.$$
$$\left.-\frac{\Gamma(\kappa)}{\Gamma(1+\kappa)}\,_2F_1\left(\kappa+1,\kappa,\kappa+1,\left(\frac{X_{0,\kappa+1}}{2U}\right)\kappa\right)\right\}+\sqrt{\sigma}\left(1-2\sigma^{-1}Z_iUM_{0,0,\kappa+1}\right)$$
$$-\varepsilon\beta\sqrt{\gamma}\frac{Z_n}{Z_i}\left(1+2\gamma^{-1}Z_nUM_{0,0,\kappa+1}\right)=0\ ,\qquad(13)$$

where we denote that

$$\frac{\rho_{0,0,\kappa+1}}{C_{0,0,\kappa+1}}=\frac{2\Gamma(\kappa+1)}{\kappa^2\Gamma(\kappa-1)},\ X_{0,\kappa+1}=\frac{2\Gamma\left(\kappa-\frac{1}{2}\right)}{\kappa\,\Gamma\left(\kappa-\frac{3}{2}\right)},\ D_{0,0,\kappa+1}=\frac{\Gamma(\kappa)}{\kappa\,\Gamma(\kappa-1)},\ \text{and}\ M_{0,0,\kappa+1}=\frac{\Gamma(\kappa)\Gamma\left(\kappa-\frac{3}{2}\right)}{2\Gamma(\kappa-1)\Gamma\left(\kappa-\frac{1}{2}\right)}.$$

After the hyper-geometric functions are calculated, Eq.(13) can be further simplified to the following equation, which is somewhat different from that in Ref.[25] because we used Eq.(7) for the negative ions, i.e., Eq.(13) becomes

$$-\mu\left(1-\frac{Z_n}{Z_i}\varepsilon+Z_iPU\right)\left(1-\frac{2U}{2\kappa-3}\right)^{-\kappa+1}+\sqrt{\sigma}\left(1-\frac{Z_iU}{\sigma}\frac{2(\kappa-1)}{2\kappa-3}\right)-\varepsilon\beta\sqrt{\gamma}\frac{Z_n}{Z_i}\left(1+\frac{Z_nU}{\gamma}\frac{2(\kappa-1)}{2\kappa-3}\right)=0.\quad(14)$$

When we take the three parameters as $\alpha = r = 0$ and $q \to \infty$, Eq. (12) can become the case for the same dusty plasma with a Maxwellian velocity distribution[18]. Namely, Eq.(12) becomes

$$-\mu\left(1-\frac{Z_n}{Z_i}\varepsilon+Z_iPU\right)\exp(U)+\sqrt{\sigma}\left(1-\frac{Z_iU}{\sigma}\right)-\varepsilon\beta\sqrt{\gamma}\frac{Z_n}{Z_i}\left(1+\frac{Z_nU}{\gamma}\right)=0\ .\qquad(15)$$

**Numerical analysis of the dust surface potential.** − In order to study further the relationship between the normalized dust surface potential $U$ and the normalized dust number density $P$ in the dusty plasma as well as its dependence on the three parameters ($\alpha$, $r$, $q$) in the non-Maxwellian velocity distribution, as an example, we make the numerical analyses of the relationship based on Eq.(12). The relationship between $U$ and $P$ in Eq.(12) has been shown for the different values of the parameters, $\alpha$, $r$ and $q$, respectively, in the Fig. 1-3. And at the same time, for comparison with the same dusty plasma with the kappa-distribution and with the Maxwellian distribution, the relationships between $U$ and $P$ in Eq.(14) for the kappa-distribution at $\kappa = 29$ and in Eq.(15) for the Maxwellian distribution are also plotted in fig. 1-3. The dependences of the relationship between $U$ and $P$ on those plasma quantities related to the negative ions such as $\varepsilon$, $\gamma$ and $Z_n$ are shown, respectively, in fig. 4-6.

In all these figures, the vertical axis is taken Log$P$, the transverse axis is taken $U$, and value range of the plasma quantities are taken consistent with the low temperature laboratory plasma [18-20,23] as $\varepsilon = 0\sim0.8$, $\gamma = 0.1\sim1$, $\mu = 242.8$, $\sigma = \beta = Z_i = 1$, and $Z_n = 1,2$.

In Fig. 1, we show the relation between the dust surface potential $U$ and the noamalized dust density $P$ for three different values of $\alpha$ when the parameters $r$ and $q$ are fixed as $r=3$ and $q=30$. The red dotted line is the relation for the kappa-distribution at $\kappa=29$. We can see that the differences are evident when $U$ is relatively smaller (e.g., $U \tilde{<} -2.2$). With increase of $U$ (e.g., $U \tilde{>} -2.2$), the differences in $U$ will weaken soon and $U$ become the same one gradually. For the fixed $r$ and $q$, the dust surface potential $U$ decreases (the absolute value increases) as the parameter $\alpha$ increases and they are generally larger than that for the kappa-distribution and the Maxwellian distribution in the same dust plasma.

In Fig. 2. we show the relation between the dust surface potential $U$ and the noamalized dust density $P$ for three different values of $r$ when the parameters $\alpha$ and $q$ are fixed as $\alpha=0.06$ and $q=30$. The red dotted line is the relation for the kappa- distribution at $\kappa=29$. We can see that the differences are evident when $U$ is relatively smaller (e.g., $U <-1.9$). With increase of $U$ (e.g., $U > -1.9$), the differences will weaken soon and $U$ become the same one gradually. For the fixed $\alpha$ and $q$, the dust surface potential $U$ increases (the absolute value decreases) as the parameter $r$ increases and they are generally larger than that for the kappa-distribution and the Maxwellian distribution in the same dust plasma.

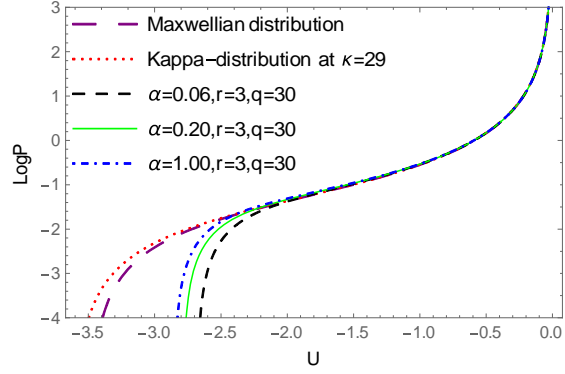

Fig. 1: The relation between $U$ and $P$ for three different values of $\alpha$ in the dusty plasma with the quantities at $\mu$=242.8, $\varepsilon$ = 0.4, $\gamma$ = 0.3, and $\sigma = \beta = Z_i = Z_n$=1.

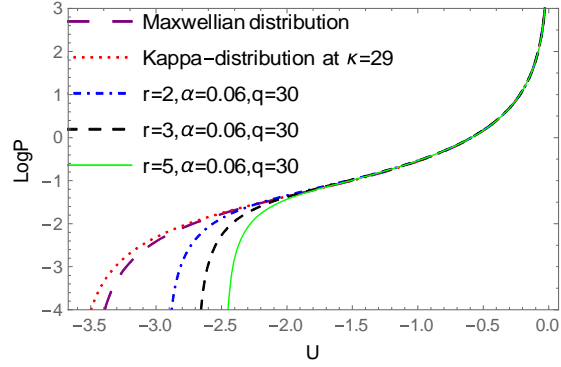

Fig. 2: The relation between $U$ and $P$ for three different values of $r$ in the dusty plasma with the quantities at $\mu$=242.8, $\varepsilon$ = 0.4, $\gamma$ = 0.3, and $\sigma = \beta = Z_i = Z_n$=1.

In Fig. 3, we show the relation between the dust surface potential $U$ and the normalized dust density $P$ for three different values of $q$ when the parameters $\alpha$ and $r$ are fixed as $\alpha$=0.06 and $r$=3. The red dotted line is also the relation for the kappa- distribution at $\kappa$=29. We can see that the differences are very evident when $U$ is relatively smaller (e.g., $U \tilde{<} -2.0$). With increase of $U$ (e.g., $U > \tilde{-}2.0$), the differences will weaken soon and $U$ become the same one gradually. For the fixed $\alpha$ and $r$, the dust surface potential $U$ increases (the absolute value decreases) as the parameter $q$ increases and they are generally larger than that for the kappa-distribution and the Maxwellian distribution in the same dusty plasma.

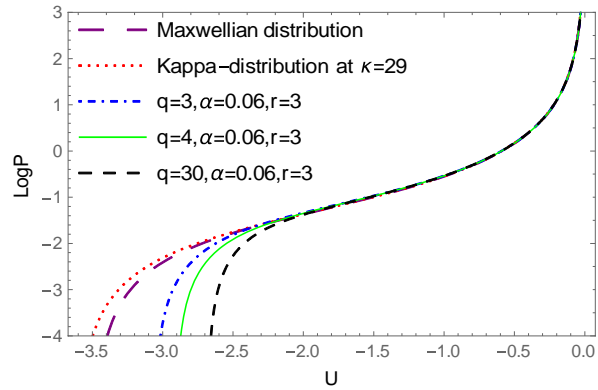

Fig. 3: The relation between $U$ and $P$ for three different values of $q$ in the dusty plasma with the quantities at $\mu$=242.8, $\varepsilon$ = 0.4, $\gamma$ = 0.3, and $\sigma = \beta = Z_i = Z_n$=1.

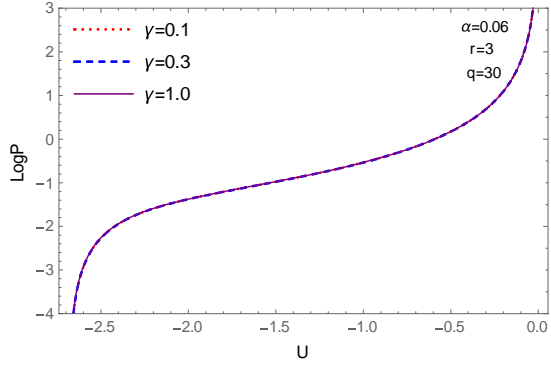

Fig. 4: The relation between $U$ and $P$ for three different values of $\gamma$ in the dusty plasma with the quantities at $\mu$=242.8, $\varepsilon$= 0.4, and $\sigma=\beta=Z_i=Z_n$=1.

In Fig. 4, we show the relation between the dust surface potential $U$ and the normalized dust density $P$ for three different values of $\gamma$ (a rate of the temperature of negative ions to the temperature of electrons) in the dusty plasma with the three-parameter non-Maxwell velocity distribution at $\alpha$=0.06, $r$=3 and $q$=30. We can see that the relation between $U$ and $P$ is basically irrelevant to the parameter $\gamma$ in this case, showing the effect of the change in $\gamma$ on the dust surface potential $U$ to be very small.

In Fig. 5, we show the relation between the dust surface potential $U$ and the noamalized dust density $P$ for three different values of $\varepsilon$ (the rate of density of the negative ions to density of the ions) in the dusty plasma with the three-parameter non-Maxwell velocity distribution at $\alpha$=0.06, $r$=3 and $q$=30. We can see that the dust surface potential $U$ increases (the absolute value decreases) generally with increase of $\varepsilon$, this is because when the rate $\varepsilon$ increases, the number of negative ions increases which means a reduction of electrons based on the quasi-neutrality condition, therefore the negative dusty surface potential will increase(the absolute value decreases). But, with the increase of $U$, the differences in $U$ for different values of $\varepsilon$ will decrease gradually.

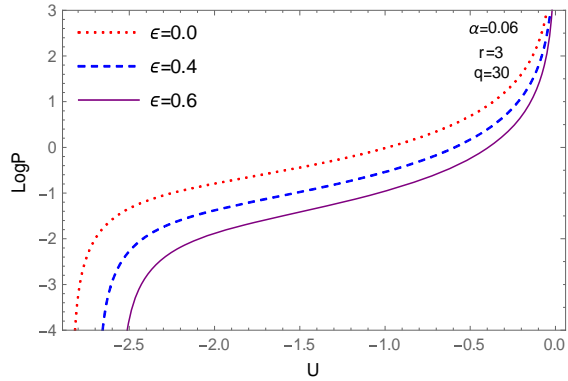

Fig. 5: The relation between $U$ and $P$ for three different values of $\varepsilon$ in the dusty plasma with the quantities at $\mu$=242.8, $\gamma$= 0.3, and $\sigma=\beta=Z_i=Z_n$=1.

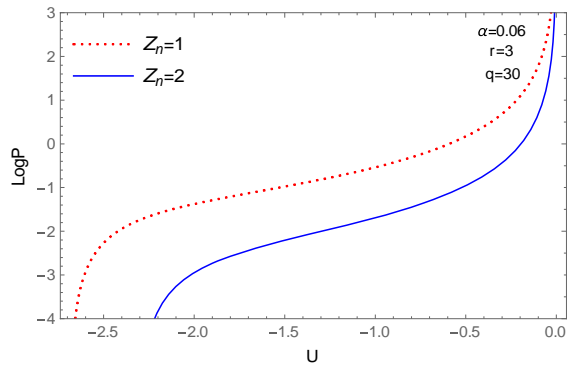

Fig. 6: The relation between $U$ and $P$ for three different values of $Z_n$ in the dusty plasma with the quantities at $\mu$=242.8, $\gamma$= 0.3, $\varepsilon$=0.4 and $\sigma=\beta=Z_i$=1.

In Fig. 6, we show the relation between the dust surface potential $U$ and the noamalized dust density $P$ for three different values of $Z_n$ (the charge amount of the negative ions) in the dusty plasma with the three-parameter non-Maxwell velocity distribution at $\alpha=0.06$, $r=3$ and $q=30$. We can see that $U$ increases (the absolute value decreases) generally with the increase of $Z_n$. This is because the increase in the charge amount of negative ions means a reduction of electrons based on the quasi-neutrality condition. So the negative dusty surface potential will increase (the absolute value decreases), but with the increase of $U$, the differences in $U$ for different values of $Z_n$ will decrease gradually.

In all the analyses of these figures, we conclude that the normalized dust surface potential $U$ increases (the absolute value of $U$ decreases) monotonously with increase of the normalized number density $P$ of dust particles. And the dust surface potential $U$ in the complex dusty plasma with the three-parameter non-Maxwell velocity distribution is generally larger than (the absolute value of $U$ is generally less than) that in the same dusty plasma with the kappa-distribution and the Maxwellian distribution.

**Summary and conclusion.** – In summary, we have studied the dust surface potential of dust particles in the complex dusty plasma with electrons, positive ions, negative ions and dust particles, and they follow the three-parameter ($\alpha$, $r$, $q$) non-Maxwell velocity distribution. Based on the orbital limit theory, we derived the current expressions of electrons, positive ions and negative ions, respectively. When the dust particles reach the electrostatic equilibrium, that is the charge: $Q_d$ = const., we get the current balance equation. Using this equation as well as the quasi-neutral condition, the relationship between the normalized dust surface potential $U$ and the normalized dust number density $P$ in the dusty plasma is derived for the kinetic non-thermal negative ions, which are given by Eq.(12). The relationship depends on the three parameters ($\alpha$, $r$, $q$) in the non-Maxwell velocity distribution. And when we take $\alpha = r = 0$ and $q = \kappa +1$, or $\alpha = r = 0$ and $q = \infty$, Eq.(12) can perfectly recover the relationship for the same dusty plasma with the kappa-distribution and that with the Maxwellian velocity distribution, respectively.

In order to analyze the dependence of the relationship between $U$ and $P$ on the three parameters ($\alpha$, $r$, $q$) in the non-Maxwell distribution more clearly, as an example we make the numerical analyses of the relationship in Eq.(12) for the three parameters ($\alpha$, $r$, $q$). The results are illustrated by Figures 1-3, respectively. And for comparison, in the Figures 1-3 we also plot the relationship in Eq.(14) for the kappa-distribution and in Eq.(15) for the Maxwellian velocity distribution, respectively, in the same dusty plasma.

The dependence of the relationship in Eq.(12) between $U$ and $P$ on the other plasma quantities such as the temperature ratio $\gamma$ of negative ions to electrons, the density ratio $\varepsilon$ of negative ions to positive ions, the charge number $Z_n$ of negative ions are numerically analyzed and illustrated in Figures 4-6, respectively.

In conclusion, we have shown that the normalized dust surface potential $U$ is negative and increases (the absolute value decreases) monotonously with increase of the normalized number density $P$ of dust particles. And the dust surface potential $U$ in the complex dusty plasma with the three-parameter non-Maxwell velocity distribution is generally larger than (the absolute value of $U$ is generally less than) that in the same plasma with the kappa-distribution and the Maxwellian distribution.

The dependence of the dust surface potential $U$ on the three parameters ($\alpha$, $r$, $q$) are very evident in the present dusty plasma when $U$ is relatively smaller, but with increase of $U$ (the absolute value of $U$ decreases ), such dependence will weaken soon and $U$ become independent of the three parameters gradually. Specifically, when the dust surface potential $U$ is relatively smaller, $U$ will decrease (the absolute value increases) with increase of the parameter $\alpha$, but $U$ will increase (the absolute value decreases) with increase of the other parameters $r$ and $q$, respectively.

The dependence of the dust surface potential $U$ on the temperature ratio $\gamma$ of negative ions to electrons in the present dusty plasma is very small, so the relation between $U$ and $P$ is basically irrelevant to the parameter $\gamma$ in this situation of the dusty plasma. The dependence of the dust surface potential $U$ on the density ratio $\varepsilon$ of negative ions to positive ions and on the charge number $Z_n$ of negative ions are both evident for all $U<0$ in the present dusty plasma, and $U$ will increases (the absolute value decrease) with increase of the parameters $\varepsilon$ and $Z_n$, respectively.

At the end, we study the dust charging model of the complex dusty plasma with a new and universal three-parameter non-Maxwell distribution, and derive the property of the currents of electrons and ions (positive ions/negative ions) as well as the dust surface potential. The dusty plasma studied here contains the most extensive non-Maxwell velocity distributions of the complex plasmas, in certain cases of the parameters, such as the two-parameter distribution, the Vasyliunas-Cairns distribution, the power-law $q$-distribution in nonextensive statistics, the Cairns-distribution, the kappa-distribution, and also the Maxwellian distribution etc. Therefore the study has great application potential and will be of broad interest in many fields including plasma physics.

\*\*\*

This work was supported by the National Natural Science Foundation of China under Grant No. 11775156.

## Appendix

The normalization of the distribution function is made simply by the equation,

$$4\pi Y_j \int_0^\infty \left(1+\alpha \frac{V_j^4}{V_{Tj}^4}\right)\left[1+\frac{1}{q-1}\left(\frac{V_j^2}{X_{r,q}V_{Tj}^2}\right)^{r+1}\right]^{-q} V_j^2 dV_j = N_j, \tag{A.1}$$

where

$$\int_0^\infty \left(1+\alpha \frac{V_j^4}{V_{Tj}^4}\right)\left[1+\frac{1}{q-1}\left(\frac{V_j^2}{X_{r,q}V_{Tj}^2}\right)^{r+1}\right]^{-q} V_j^2 dV_j$$

$$= \frac{X_{r,q}^{\frac{7}{2}}V_{Tj}^3(q-1)^{\frac{7}{2r+2}}}{21\Gamma(q)}\left[3\alpha\Gamma\left(q-\frac{7}{2r+2}\right)\Gamma\left(1+\frac{7}{2r+2}\right)+7\Gamma\left(q-\frac{3}{2r+2}\right)\Gamma\left(1+\frac{3}{2r+2}\right)\frac{X_{r,q}^{-2}}{(q-1)^{\frac{4}{2r+2}}}\right]$$

$$= \frac{X_{r,q}^{\frac{3}{2}}V_{Tj}^3}{3\Gamma(q)}(q-1)^{\frac{3}{2r+2}}\Gamma\left(q-\frac{3}{2r+2}\right)\Gamma\left(1+\frac{3}{2r+2}\right)\left[\alpha X_{r,q}^2 \frac{(q-1)^{\frac{2}{r+1}}\Gamma\left(q-\frac{7}{2r+2}\right)\Gamma\left(\frac{7}{2r+2}\right)}{\Gamma\left(q-\frac{3}{2r+2}\right)\Gamma\left(\frac{3}{2r+2}\right)}+1\right]$$

$$= \frac{X_{r,q}^{\frac{3}{2}}V_{Tj}^3}{3\Gamma(q)}(9\alpha\eta_{r,q}+1)(q-1)^{\frac{3}{2r+2}}\Gamma\left(q-\frac{3}{2r+2}\right)\Gamma\left(1+\frac{3}{2r+2}\right).$$

We therefore have that

$$Y_j = \frac{3N_j}{4\pi V_{Tj}^3} \frac{X_{r,q}^{-\frac{3}{2}}\Gamma(q)(q-1)^{-\frac{3}{2(1+r)}}}{(1+9\alpha\eta_{r,q})\Gamma\left(q-\frac{3}{2(1+r)}\right)\Gamma\left(1+\frac{3}{2(1+r)}\right)} \equiv \frac{3N_j}{4\pi V_{Tj}^3}\frac{\rho_{\alpha,r,q}}{X_{r,q}^{\frac{3}{2}}}. \tag{A.2}$$

The electron current is derived by

$$I_e = -4\pi e \int_{\sqrt{2Q_e\phi_d/m_e}}^\infty V_e^3 \sigma_e^d f_e(V_e) dV_e$$

$$= -4e\pi^2 r_d^2 Y_e \int_{\sqrt{2Q_e\phi_d/m_e}}^\infty V_e^3 \left(1+\frac{2e\phi_d}{m_e V_e^2}\right)\left(1+\alpha\frac{V_e^4 m_e^2}{T_e^2}\right)\left[1+\frac{1}{q-1}\left(\frac{V_e^2 m_e}{X_{r,q}T_e}\right)^{r+1}\right]^{-q} dV_e$$

$$= -3e\pi N_e r_d^2 \frac{\rho_{\alpha,r,q}}{V_{Te}^3 X_{r,q}^{3/2}}\left\{\frac{8\alpha e^4 \phi_d^4 (q-1)^q}{T_e^2 m_e^2}\left(\frac{T_e X_{r,q}}{-2e\phi_d}\right)^{q(1+r)}\left[\frac{H_1}{q+qr-4}-\frac{H_2}{q+qr-3}\right]\right.$$

$$\left.+\frac{2e^2\phi_d^2(q-1)^q}{m_e^2}\left(\frac{T_e X_{r,q}}{-2e\phi_d}\right)^{q(1+r)}\left[\frac{H_3}{q+qr-2}-\frac{H_4\Gamma\left(q-\frac{1}{1+r}\right)}{(1+r)\Gamma\left(\frac{r}{1+r}+q\right)}\right]\right\}$$

$$= 3\pi eN_e r_d^2 \vartheta_e \left\{\frac{8\alpha\rho_{\alpha,r,q}U^4(q-1)^q}{X_{r,q}^2}\left(\frac{X_{r,q}}{-2U}\right)^{q(1+r)}\left(\frac{H_1}{q+qr-4}-\frac{H_2}{q+qr-3}\right)\right.$$

$$\left.+\frac{2\rho_{\alpha,r,q}U^2(q-1)^q}{X_{r,q}^2}\left(\frac{X_{r,q}}{-2U}\right)^{q(1+r)}\left[\frac{H_3}{q+qr-2}-\frac{H_4\Gamma\left(q-\frac{1}{1+r}\right)}{(1+r)\Gamma\left(\frac{r}{1+r}+q\right)}\right]\right\}$$

$$= -6\pi eN_e r_d^2 \vartheta_e \left(\frac{X_{r,q}}{-2U}\right)^{q(1+r)}\frac{(q-1)^q \rho_{\alpha,r,q} U^2}{X_{r,q}^2}\left[4\alpha U^2\left(\frac{H_1}{q+qr-4}-\frac{H_2}{q+qr-3}\right)\right.$$

$$\left.+\frac{H_3}{q+qr-2}-\frac{H_4\Gamma\left(q-\frac{1}{1+r}\right)}{(1+r)\Gamma\left(\frac{r}{1+r}+q\right)}\right]. \tag{A.3}$$

The positive ion current is derived by

$$I_i = 4\pi Z_i e \int_0^\infty V_i^3 \sigma_i^d f_i(V_i) dV_i$$

$$= 4\pi^2 r_d^2 Y_i Z_i e \int_0^\infty V_i^3 \left(1-\frac{2Z_i e\phi_d}{mV_i^2}\right)\left(1+\alpha\frac{V_i^4 m_i^2}{T_i^2}\right)\left[1+\frac{1}{q-1}\left(\frac{V_i^2 m_i}{X_{r,q}T_i}\right)^{r+1}\right]^{-q} dV_i$$

$$= 4\pi^2 r_d^2 Z_i e Y_i \left[ \frac{\alpha T_i^2 X_{r,q}^4}{8m_i^2 \Gamma(q)} (q-1)^{\frac{4}{1+r}} \Gamma\left(\frac{5+r}{1+r}\right)\Gamma\left(q-\frac{4}{1+r}\right) \right.$$

$$- \frac{\alpha Z_i e\phi_d T_i X_{r,q}^3}{3m_i^2 \Gamma(q)} (q-1)^{\frac{3}{1+r}} \Gamma\left(\frac{4+r}{1+r}\right)\Gamma\left(q-\frac{3}{1+r}\right) + \frac{T_i^2 X_{r,q}^2}{4m_i^2 \Gamma(q)} (q-1)^{\frac{2}{1+r}} \Gamma\left(\frac{3+r}{1+r}\right)\Gamma\left(q-\frac{2}{1+r}\right)$$

$$\left. - \frac{Z_i e\phi_d T_i X_{r,q}}{m_i^2 \Gamma(q)} (q-1)^{\frac{1}{1+r}} \Gamma\left(\frac{2+r}{1+r}\right)\Gamma\left(q-\frac{1}{1+r}\right) \right]$$

$$= 3\pi r_d^2 N_i Z_i e \frac{\rho_{\alpha,r,q}}{X_{r,q}^{3/2}} \left[ \frac{\alpha X_{r,q}^4 V_{Ti}}{8\Gamma(q)} (q-1)^{\frac{4}{1+r}} \Gamma\left(\frac{5+r}{1+r}\right)\Gamma\left(q-\frac{4}{1+r}\right) \right.$$

$$- \frac{\alpha Z_i e\phi_d X_{r,q}^3 V_{Ti}}{3\Gamma(q)T_i} (q-1)^{\frac{3}{1+r}} \Gamma\left(\frac{4+r}{1+r}\right)\Gamma\left(q-\frac{3}{1+r}\right) + \frac{X_{r,q}^2 V_{Ti}}{4\Gamma(q)} (q-1)^{\frac{2}{1+r}} \Gamma\left(\frac{3+r}{1+r}\right)\Gamma\left(q-\frac{2}{1+r}\right)$$

$$\left. - \frac{Z_i e\phi_d X_{r,q} V_{Ti}}{T_i \Gamma(q)} (q-1)^{\frac{1}{1+r}} \Gamma\left(\frac{2+r}{1+r}\right)\Gamma\left(q-\frac{1}{1+r}\right) \right]$$

$$= \frac{3\pi r_d^2 N_i Z_i e \vartheta_i (q-1)^{\frac{1}{2(r+1)}}}{(1+9\alpha\eta_{r,q})\Gamma\left(1+\frac{3}{2(1+r)}\right)} \left[ \frac{(q-1)^{\frac{4}{2(1+r)}} \Gamma\left(1+\frac{4}{1+r}\right)\Gamma\left(q-\frac{4}{1+r}\right)}{\Gamma\left(q-\frac{3}{2(1+r)}\right)} \left( \frac{\alpha X_{r,q}^2}{8} - \frac{\alpha Z_i e\phi_d X_{r,q}^2}{3m_i \vartheta_i^2} \frac{\Gamma\left(1+\frac{3}{1+r}\right)\Gamma\left(q-\frac{3}{1+r}\right)}{(q-1)^{\frac{1}{1+r}} \Gamma\left(1+\frac{4}{1+r}\right)\Gamma\left(q-\frac{4}{1+r}\right)} \right) \right.$$

$$\left. + \frac{\Gamma\left(1+\frac{2}{1+r}\right)\Gamma\left(q-\frac{2}{1+r}\right)}{4\Gamma\left(q-\frac{3}{2(1+r)}\right)} \left( 1 - \frac{2Z_i e\phi_d}{m_i \vartheta_i^2} \frac{2\Gamma\left(1+\frac{1}{1+r}\right)\Gamma\left(q-\frac{1}{1+r}\right)}{(q-1)^{\frac{1}{1+r}} \Gamma\left(1+\frac{2}{1+r}\right)\Gamma\left(q-\frac{2}{1+r}\right)} \right) \right]$$

$$= 3\pi r_d^2 \vartheta_i N_i Z_i e \left[ A_{\alpha,r,q} \alpha X_{r,q}^2 \left(\frac{1}{8} - \frac{Z_i e\phi_d}{3m_i \vartheta_i^2} B_{\alpha,r,q}\right) + C_{\alpha,r,q}\left(1 - \frac{2Z_i e\phi_d}{m_i \vartheta_i^2} D_{\alpha,r,q}\right) \right]. \tag{A.4}$$

The negative ion current is derived by

$$I_n = -4\pi Z_n e \int_{\sqrt{2Q_n \phi_d/m_n}}^{\infty} V_n^3 \sigma_n^d f_n(V_n) dV_n,$$

$$= -4Z_n e\pi^2 r_d^2 Y_n \int_{\sqrt{2Q_n \phi_d/m_n}}^{\infty} V_n^3 \left(1+\frac{2Z_n e\phi_d}{m_n V_n^2}\right)\left(1+\alpha\frac{V_n^4 m_n^2}{T_n^2}\right)\left[1+\frac{1}{q-1}\left(\frac{V_n^2 m_n}{X_{r,q} T_n}\right)^{r+1}\right]^{-q} dV_n$$

$$= -3Z_n e\pi N_n r_d^2 \frac{\rho_{\alpha,r,q}}{V_{Tn}^3 X_{r,q}^{3/2}} \left[ \frac{8\alpha Z_n^4 e^4 \phi_d^4 (q-1)^q}{T_n^2 m_n^2} \left(\frac{T_n X_{r,q}}{-2Z_n e\phi_d}\right)^{q(1+r)} \left(\frac{h_1}{q+qr-4} - \frac{h_2}{q+qr-3}\right) \right.$$

$$\left. + \frac{2Z_n^2 e^2 \phi_d^2 (-1+q)^q}{m_n^2} \left(\frac{T_n X_{r,q}}{-2Z_n e\phi_d}\right)^{q(1+r)} \left(\frac{h_3}{q+qr-2} - \frac{h_4 \Gamma\left(q-\frac{1}{1+r}\right)}{(1+r)\Gamma\left(\frac{r}{1+r}+q\right)}\right) \right]$$

$$= -6\pi Z_n e N_n r_d^2 \vartheta_n \left(\frac{\gamma X_{r,q}}{-2Z_n U}\right)^{q(1+r)} \frac{(q-1)^q Z_n^2 U^2 \rho_{\alpha,r,q}}{\gamma^2 X_{r,q}^2} \left[ \frac{4Z_n^2 \alpha U^2}{\gamma^2} \left(\frac{h_1}{q+qr-4} - \frac{h_2}{q+qr-3}\right) \right.$$

$$\left. + \frac{h_3}{q+qr-2} - \frac{h_4 \Gamma\left(q-\frac{1}{1+r}\right)}{(1+r)\Gamma\left(\frac{r}{1+r}+q\right)} \right]. \tag{A.5}$$